\documentclass[12pt,preprint]{article}
\usepackage{epsfig,amsmath}

\renewcommand{\eqref}[1]{(\ref{#1})}

\begin{document}
\begin{titlepage}
\mbox{} \vskip3cm \centerline{\Large\bf \sf Ab Initio Modeling of
Ecosystems with Artificial Life}
\vskip 0.5cm
\vskip 0.25in
\centerline{\large \bf \sf C. Adami$^{1,2}$}
\vskip 0.25in

\centerline{\it $^1$Jet Propulsion Laboratory 126-347, California
Institute of Technology, Pasadena, CA 91109}
\vskip 0.25cm
 \centerline{\it $^2$Digital Life
Laboratory 136-93}\centerline{\it California
Institute of Technology, Pasadena, CA 91125}
\vskip 2cm

\vskip 2cm


\abstract{Artificial Life provides the opportunity to study the emergence and
evolution of simple ecosystems in real time. We give an overview of
the advantages and limitations of such an approach, as well as its
relation to individual-based modeling techniques. The Digital Life
system {\rm Avida} is introduced and prospects for experiments with
{\it ab initio} evolution (evolution ``from scratch''), maintenance, as well as
stability of ecosystems are discussed.}
\end{titlepage}

\vskip 0.25cm

\section{Introduction}
Individual-based ecological modeling seems to be an unstoppable trend
in modern ecological science, but it is not without its
problems~\cite{Grimm00}. The attraction of this approach stems from
the promise that individual-based models (IBMs) might capture emergent
effects on a macroscopic level while only implementing simple
interactions between agents on the microscopic level. This bottom-up
approach is, in fact, a legacy of the early years of Artificial Life
(see, e.g.,~\cite{alife2}) and has probably produced as much success
as failure. The problems are not difficult to discern. On the
microscopic level, the salient characteristics of the agents have to
be decided upon, and it is rather rare that such decisions are made
with strict theoretical concepts in mind. More often than not, they
are subject to change and tinkering.  On the macroscopic level,
emergent effects are (precisely because of their nature) often
subjective. The interaction between tinkering and subjective
appreciation produces, again more often than not, a tendency to select
effects via {\em parameter mutation}, i.e., a particular set of
parameters may tend to be adopted as standard simply because of the
subjectively interesting patterns it may produce.  Thus, in the
language of evolutionary biology, you often ``get what you select
for''. This modeling disease, well-known from Artificial Life, is
certainly also a problem in IBMs.

Another problem concerns
the complexity of most IBMs and the ensuing difficulty in
communicating salient, ``universal'', results. To some extent, the
modeling of complex systems defeats its purpose if the complexity
cannot be {\em reduced}. After all, ``complex'', when used together
with ``systems'', is just a synonym for ``not-understood''.

Thus, sane IBMs should strive at reducing the complexity of a
scenario, by treating cases that are as simple as possible, while retaining
the essential characteristic being addressed. Furthermore, limiting cases
should be included that have well-known theoretical solutions, such
that baselines can be established that are beyond doubt. Because
ecological modeling is supposed to replace at least some experimental
work, there is no doubt that it has to be supplemented and checked
against theory as much as possible~\cite{Grimm00}.

In this contribution, I would like to propose a modeling paradigm for
IBMs which avoids the pitfalls mentioned above. It is not a panacea,
as its applicability to the ecological sciences is severely
limited. As will become obvious, it has nothing to say about higher
animals, food webs, or predator-prey interactions. However, it carries
the promise to answer fundamental questions in ecology, such as
concerning the emergence, maintenance, and stability of simple
ecologies in simple environments. The way the model avoids the
possible problems listed above is that, in the first place, the
organisms in question---the microscopic agents---have not been
designed. Instead, they themselves emerge via adaptation to an
artificial world. Secondly, the system is simple enough that the
fundamental dynamics of these organisms is well-understood
theoretically (and by now experimentally), at least in the
single-niche environment. Thus, there is ample room for validation and
cross-checking. The organisms I am speaking about, rather than being
abstract agents, are in fact self-replicating computer programs that
live in the random access memory (RAM) of the computers within which the
virtual world is built. They are a form of life~\cite{harvey99} whose
primary purpose has been research into possible universal
characteristics of evolution in simple living systems. This modeling
platform (or more precisely, experimental platform, see below), called
{\sf avida}\footnote{{\sf Avida} is free software and can be
downloaded from the Digital Life Laboratory's site at {\tt
http://dllab.caltech.edu}}, has been used in the single-niche mode,
without coevolution, for the last seven years. The reasons for such
caution lie precisely in validation: there is no logic in facing the
complexity of multi-niche populations adapting to local conditions if
the single-niche, homogeneous, situation is not fully
understood. However, the time for the next step appears to have come.

\section{Simulation or Experiment?}
The main distinction between the kind of individual-based modeling
described above and the approach followed in Artificial Life (in
particular the type of Digital Life described below) is that the
quantities which carry the system's essential properties (the``degrees
of freedom'') under consideration in Artificial
Life, whether they be robotic, genetic, or abstract (such as cellular
automata) are given a {\em physical}, rather than mathematical,
representation. The dynamics (collective or otherwise) of these
degrees of freedom is subsequently computed and {\em observed}. This
distinction is, in most cases, not arbitrary. It relies on a
fundamental discovery which lies at the heart of computational
science, namely Turing's famous theorem concerning the universality of
computation~\cite{turing36}. Turing's theorem implies a duality
between computation and physical dynamics which has wide-ranging
consequences not only for computation, but indeed for physics. It
states that any dynamical system with physical degrees of freedom that
is ``complicated enough''\footnote{The minimum complexity requirement
is that a universal Turing machine can be constructed from these
degrees of freedom.}  is capable of universal computation, which means
that a computer constructed using these degrees of freedom can
calculate ``any''\footnote{In reality ``almost any''
function. Technically, at most partially recursive functions can be
calculated by universal computers, a restriction due to the ``Halting
Problem''.}  function. The theorem also implies that all universal
computers are strictly equivalent, in the sense that any such computer
can {\em simulate} any other. An under-appreciated consequence of the
theorem is that it works ``both ways'': If ``complex enough'' dynamics
imply universal computation, then it is clear that any universal
computer (such as, e.g., any von Neumann-architecture computer) has to
have ``interesting'' physical dynamics at its core. With
``interesting'' I mean that the degrees of freedom and their dynamics
could be used, or shaped, to do almost anything.

Let me give an example. With a computer, there are two ways to
investigate wave propagation in media. The usual one is to write down
the relevant differential equations and solve them numerically. In this case,
the wave is represented mathematically (say: sin($x$)), and we would
not be tempted to assert that there is an actual wave anywhere inside
of the computer. This is, of course, the central idea in
simulation. Here is another way to solve this problem with a
computer. Suppose that instead of programming the differential
equations, I instead program memory locations within RAM to interact
in a specific manner. In particular, I program adjacent locations
(which we can view as potential wells carrying charges) to interact
like springs with a particular mass, spring constant, and a given amount
of friction. It would then be possible to observe an excitation
propagate through the RAM of the computer precisely like the sine wave
in the simulation, only that this wave would be {\em real}. Granted,
this would be a perverse way of solving wave propagation with a
computer, but the example serves to illustrate that universal
computers do indeed rely on physical dynamics at their core, and that
these physical degrees of freedom (charges in potential wells) are as
real as billiard balls or mechanical cogs or biochemical molecules,
all of which have been used to construct universal computers.

When a computer is used in such a ``perverse'' manner, it is clear
that one is performing experiments, not simulations. Under which
circumstances does this approach present an advantage over simulation?
Precisely in the case where the construction of the physical dynamics
would be far too complicated in any other medium than the computer,
and when the architecture of the computer is best suited for the
problem at hand. Because the physical degrees of freedom in a computer
are primarily used to store, transmit, and manipulate information,
physical dynamics of information-bearing degrees of freedom are most
easily implemented. This encompasses the dynamics of simple living
systems, in particular those whose essence is informational (i.e.,
self-replicating molecules). A typical example would be a computer
virus.  Computer viruses physically populate a computer's memory and
physically replicate in it. As it is clear that the viruses are not
simulated, and as it is clear from the duality mentioned above that there
is no reason to draw a fundamental distinction between one sort of
information-bearing degree of freedom (e.g., biomolecules) or charges
in potential wells, self-replicating programs within a computer's
memory have to be considered on an equal footing as, say, {\it
Mycoplasma mycoides} thriving in our nasal passages. Turing's insight,
thus, demolishes the barrier between ``real'' and ``artificial'' life.

\section{Digital Life}
The field of Digital Life was, in fact, directly inspired by computer
viruses. In 1989, Steen Rasmussen at Los Alamos National Laboratory
created a ``reserve'' for computer viruses inside of a
computer~\cite{RAS1}, by creating virtual CPUs that executed programs
written in {\sf Redcode}, a type of assembly language used in a
computer game called ``CoreWars'', where the objective is to write
code that invades and takes over another computer's memory. It turned
out that the best strategy to win this game was to write
self-replicating programs, and these were used by Rasmussen to
inoculate that special space inside of the computer.  As a
consequence, the ``world'' in which these programs live is virtual,
i.e., simulated, while the actual programs (because they are physical)
are real.  This can be roughly compared to what happens in {\it in
vitro} experiments with {\it E. coli} bacteria. The environment,
(namely the Petri dish with its nutrients) that these bacteria live in
is entirely artificial, and very controlled. The bacteria on the other
hand are real. While leading to important insights, Rasmussen's
experiments in evolution were not successful because his choice of
world led to very fragile programs: while trying to copy themselves
they inevitably overwrote adjacent programs, such that populations
died very quickly. This shortcoming was lifted by Tom Ray with the now
famous {\sf tierra} software~\cite{RAY1}, in which the first
interesting dynamics of ``alien life'' could be observed. Ray's {\sf
tierra} inspired the creation of the {\sf avida} system at Caltech in
1993~\cite{AB1,adami98}; {\sf avida} is the most widely used Digital Life
platform today. Because of the flexibility of {\sf avida} as an
experimental platform, it has been used to address a wide variety of
problems in the evolution and dynamics of simple living
systems~\cite{CHU1,adami98a,ofria99a,ofria99b,Lenskietal99,adami00,wagenaar00}.

\begin{figure}[tb]
  \centerline{ \epsfig{file={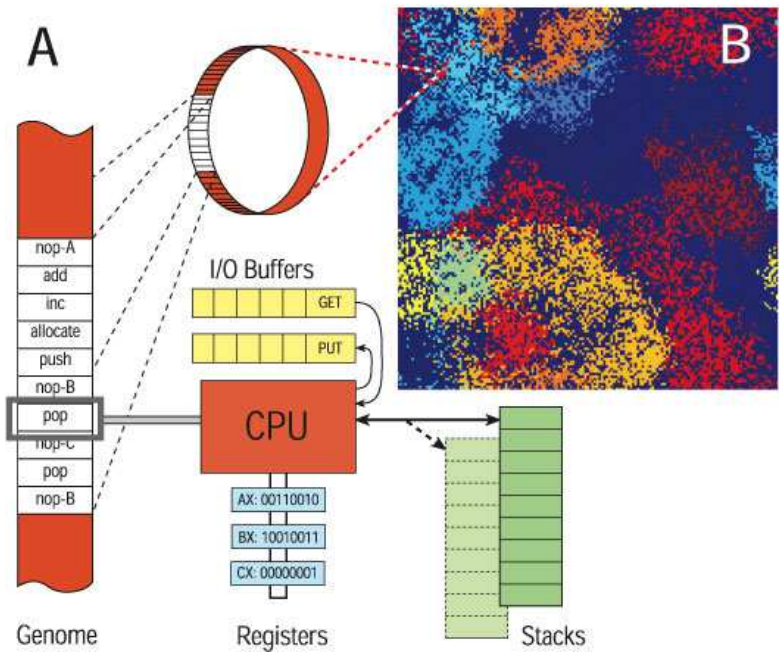},width=.75\columnwidth} }
\caption{The virtual world of avidians. ({\bf A}) Simple virtual
CPU that executes avidian code. The CPU
  operates on three registers (blue) and two stacks (green). Input and
  output from and to the environment is achieved via dedicated I/O
  buffers (yellow). ({\bf B}) Physical arrangement of programs on a 2D
  lattice in the {\sf Avida} world. Different colors indicate different
  program genotypes.
\label{fig:world}}
\end{figure}

From its initial design, {\sf avida} emphasized simplicity and
accountability. As an experimental platform, it would have to be
possible to focus on particular aspects of living systems while
``turning off'' those that could detract from the question being
asked. Thus, for example, sexual crossover between organisms was not
to be implemented before the dynamics of asexual reproduction was
fully understood and mapped out (a stage yet to be reached). World
geometry was chosen either as a two-dimensional grid that wraps on
itself (a flat torus) in order to avoid boundary effects, or else a
well-stirred reactor without any geometry. In both cases, the world is
isotropic and homogeneous, i.e., it is a single-niche environment (see
Fig.~\ref{fig:world}).

Avidians are thus computer programs that self-replicate within a
virtual world created for them inside of a computer. Unlike ordinary
computer viruses, however, their replication mechanism is imperfect,
leading to mutations which allow them to adapt to their world and grow
in complexity.  Their genetic code consists of instructions taken from
a set designed to be simple and universal (in the sense that, again,
almost any program can be written within that language). Naturally,
many different such sets can be designed and indeed, because
different instruction sets can be viewed as different types of {\em
chemistry}, this opens up the opportunity to investigate how the
dynamics of simple living systems depends on such a choice. In
biochemistry, the equivalent would be the freedom to investigate
biochemistries based on widely different sets of amino acids. In {\sf
avida}, we usually use a standard instruction set of 28 different
instructions which are superficially similar to Intel i860 assembly
instructions, liberally supplemented with instructions that allow
self-replication. Examples of such instructions are {\tt copy}, which
copies an instruction from one memory location to another (and which
has a probability to fail set by the mutation rate), {\tt allocate},
which allocates memory space before replication can begin, and {\tt
divide}, which separates a mother and daughter program and places the
daughter, either near the progenitor on the grid or anywhere, depending on
the choice of geometry. Other instructions manipulate the virtual
CPU's registers and stacks, and perform logical, bitwise, operations
on random numbers that are abundant in the organism's
environment. Different instruction sets (artificial chemistries) can
affect the dynamics of the population markedly, in particular with
respect to evolvability and robustness~\cite{ieee2001}.

The heart of the avidian system is the energy metabolism of the
organisms. The primary resource, without which no program can survive,
is CPU time. It plays the role that the carbon source plays for
bacteria in a Petri dish, except that it is not substitutable. CPU
time is distributed in ``time slices'' to each organism in the
population. The relative amount received by each organism depends on a
number of factors. A default amount is distributed according to genome
length, in order to make the replication process genome-length
independent (the generation time is proportional to length). On top of
that, CPU ``bonus'' time is given out for those programs that have
developed computational genes. Such genes are stretches of program
which read numbers from the environment, perform computations on them,
and write them back out. Rewarded computations are, currently, logical
operations on binary numbers, with up to three inputs (this is a total
of 78 different operations). Since the only instruction available for
such computations is a logical NAND\footnote{NAND is the logical
operation which is the negation of the AND operation performed on two
binary inputs. It can be used as a primitive to construct all possible
logical operations.}, complex computations require the evolution of
significantly complex code. It is worth noting that most evolutionary
experiments are started with a simple ancestral genotype that is only
20 instructions long and whose only function is self-replication, yet
organisms with sequence lengths of several hundred instructions
performing a good fraction of all possible computations readily emerge
in these experiments. Because it is these computations which provide
the organism with the ``energy'' (in the form of CPU time) it needs to
replicate, we can think of this computational code as the genes that
code for the organism's metabolism. To this extent, we are able to
observe the emergence of metabolic genes in self-replicating
organisms, and thus the evolution of complexity~\cite{adami00}.

From a biological point of view, avidian populations are extremely
simple, by design. In particular, it is possible in this scenario to
make a clean distinction between the population and the environment
that the population is adapting to, due to the fact that there is
virtually no co-evolution in this system. On the other hand, many of
the most interesting questions in evolutionary biology, and in ecology
in particular, have to do with co-evolution of organisms and adaptation
to local conditions, for example to depletable and substitutable local
resources. In the following, I shall outline the type of changes that
in the avidian system required to investigate such
questions, and discuss possible experiments in which the emergence,
maintenance, and stability of simple ecosystems can be studied.

\section{Resource Competition and Ecosystems in Avida}
As described above, most of the adaptive activity in {\sf avida} is
geared towards the evolution of ``metabolic'', i.e., computational,
genes. Typical examples of computations (``tasks'') being rewarded and
performed would (for two inputs $i$ and $j$) be\footnote{Besides the
computational primitive NAND mentioned above, there are several other
distinct logical operations on two input bits, such as AND, OR, and
XOR. The latter, ``exclusive OR'', returns ``true'' (or the bit value
``1'') only if one and only one of the two input bits are ``true'' (or
``1'' ). XOR is an example of a difficult logical operation to perform
armed with only NAND, and a skilled programmer needs at least 19 lines
of avidian code to construct it. Nature, i.e., evolution, can do
somewhat better.}  ($i$ AND $j$), ($i$ XOR $j$), (NOT $i$ OR $j$),
etc. In a sense, these genes can be viewed as the analog of exothermal
catalytic reactions that are being carried out by the organisms, as
they allow a more ``efficient'' exploitation of the primary resource,
CPU time. The product of the reaction is the computational result,
while the substrate it uses are the numbers available in the
environment. In the standard avidian world, however, these numbers are
inexhaustible, and isotropically distributed. To move towards
exhaustible and local resources, rather than limiting the numbers you
can read, or where you can read them, each task instead is associated
with an abstract resource, whose presence is required in order for the
organisms to reap the reward of the computation. Let us imagine, for
the moment, a simple world in which there are only three different
possible tasks, say the ones introduced above. Then, we associate
resource ``A'' with AND, ``B'' with NOR\footnote{NOR stands for NOT
($i$ OR $j$), and is another independent logical operation on two input
bits.}, and ``C'' with XOR. We can
now load up the world with these resources, and we can limit them. For
example, we can stipulate that every time an organism performs an AND,
a certain amount of resource A is depleted, and similarly for the
other resources. There are a number of obvious but important
consequences of such a scenario. First of all, it is clear that
computational genes can {\em only} evolve in regions where the
corresponding resource is present. Thus, local differences in resource
abundances will lead to different genes evolving in different areas,
and a multi-niche population must form. Second, depletion of resources
forces new selective pressures on the population, and strategies must
be evolved to avoid starvation. Of course, for this to be non-trivial
we must assume that there is a certain rate of substrate renewal, and
the most economic way of implementing this is via a typical flow
reactor.

At the same time, it appears reasonable to propose that performance of
a computation in the presence of the enabling resource might transform
the resource rather than use it up. In such a model for example, we could have
\begin{eqnarray}
A &\stackrel{{\tt AND}}\longrightarrow & B \;,\nonumber \\
B &\stackrel{{\tt NOR}}\longrightarrow & C \;,\nonumber \\
C &\stackrel{{\tt XOR}}\longrightarrow & A \;.\nonumber
\end{eqnarray}
Of course, in a system such as this, any type of resource chemistry
can be implemented or tested. Still, extremely simple chemistries such as
the one above already lead to a number of interesting questions and
suggest experiments whose outcome is not immediately obvious.

\begin{figure}[ht]
 \centerline{ \epsfig{file={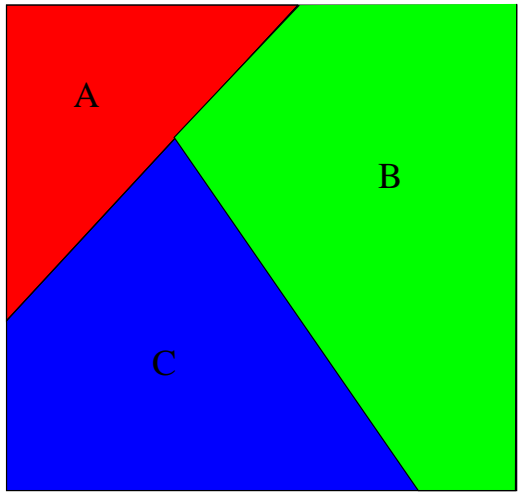},angle=-90,width=.5\columnwidth} }
\caption{Spatial distribution of transformable resources for an experiment
testing allopatric speciation, and the evolution of heterotrophy.
\label{spatial}}
\end{figure}

For example, let us imagine we start an experiment in which three
resources A, B, and C are distributed spatially such as in
Fig.~\ref{spatial}.
If resources are continuously renewed (but not transformed),
specialists will evolve in the three different habitats, and no
species can invade the other. If resources become scarce and if they
are, through usage, transformed according to the rules above, we can
expect two different scenarios. Either an organism will develop that
carries all three genes (for AND, NOR, and XOR) making it effectively
an autotroph, or we will witness colonies of heterotrophs forming at
the boundary where the three resources A, B, and C meet in
Fig.~\ref{spatial}. Such colonies would represent simple
ecosystems whose members rely on each other's metabolic products.

After such colonies are formed (and surely we can imagine breeding
much more complex ecosystems, built on more complex artificial resource
chemistries) it becomes possible to study their stability. For example,
such ecosystems can be tested at different levels of starvation. Also,
because it is possible to manipulate each species, and each resource
as well as its strength (in units ``bonus CPU time'') separately, it
ought to be possible to remove one species at a time in order to
observe either a self-healing or a collapse of the ecosystem. Such a
strategy might also serve to determine whether there are ``keystone''
species present in the ecosystem, whose removal is tantamount to
ecosystem collapse.

Another question that might be addressed experimentally with such a
system concerns the well-known``plankton paradox'' of ecology, in
which $n$ species appear to be able to coexist on $k<n$ resources,
violating the competitive exclusion principle~\cite{hutch61}. A recent
explanation of the paradox~\cite{huis00} invokes non-equilibrium
dynamics of at least three species, which can coexist on less than
three resources because three-species competition can generate
sustained oscillations. In such a scenario, species displace each
other in a cyclical fashion, dominating via usage of one resource
while becoming limited by another. While such a scenario may indeed
explain the $n>k$ puzzle, it is at this point entirely
theoretical. Within Digital Life, experiments can be carried out to
observe if, and under which circumstances, such oscillations do indeed
arise, and whether they contribute to increasing the diversity of the
population.

\section{Conclusions}
Individual-based modeling of ecosystems represents a powerful new tool
for the study of complex ecosystems, but it generates its own sets of
problems, both in the areas of validation and communication. At least
for simple ecosystems composed of simple organisms, Artificial Life
can be useful in addressing some issues fundamental to the ecological
sciences, pertaining to the emergence, evolution, maintenance, and
stability of such simple ecosystems.  The simplicity of the system,
while maintaining a complex fitness landscape, allows the design of
experiments (and their controls) with conclusive results. In such a
system, parameters can be set at will, and measurements can be
recorded noise-free that perhaps are impossible to obtain in natural
systems.  At the same time, these results have to reproducible, and
fit within a theoretical construct which allows an interpretation of
the data. Experiments, according to the standard philosophy of
Science, are the potential ``falsifiers'' of theory. Without a
framework to compare to, any amount of collected data is worthless.

While a system such as {\sf avida} has nothing to say about realistic
complex ecosystems of several trophic levels, such as are often the
object of IBMs, it is conceivable that the type of experimental
approach followed there, guided by theoretical insight and a tendency
towards parsimony, may profit IBM practitioners too.

\vskip 0.25cm
\noindent{\bf Acknowledgements}
\vskip 0.25cm

I would like to thank Steve Railsback for inviting me to speak at the
85th Annual Meeting of the Ecological Society of America (where part
of this work was presented), Roland H. Lamberson for the opportunity
to contribute to this volume, as well as an anonymous referee for
valuable comments.  This work was supported in part by the National
Science Foundation under contract No.\ DEB-9981397.  A portion of the
research described in this paper was carried out at the Jet Propulsion
Laboratory, under a contract with the National Aeronautics and Space
Administration.


\begin{thebibliography}{99}
\bibitem{Grimm00}V. Grimm. Ten years of individual-based modeling in
ecology: What have we learned and what could we learn in the future?
{\it Ecol. Model.} {\bf 115}:129--148 (1999).

\bibitem{alife2}C. G. Langton, C. Taylor, J.D. Farmer, and
S. Rasmussen, eds., {\it Artificial Life II}: Proceedings
  of the 2nd International Workshop on the Synthesis and Simulation of
  Living Systems, (Addison Wesley, Redwood City, 1992).

\bibitem{harvey99}I. Harvey. Creatures from another world. {\it Nature} {\bf
400}:618--619 (1999).

\bibitem{turing36}A. Turing, On computable numbers with an application
to the {\it Entscheidungsproblem}, {\it
Proc. Lond. Math. Soc. Ser. 2},  {\bf 43}:544, {\bf 42}:230 (1936).

\bibitem{RAS1} S. Rasmussen, C. Knudsen, R. Feldberg, and M.
Hindsholm, The Coreworld: Emergence and evolution of cooperative structures
in a computational chemistry, {\it Physica} {\bf D 42}: 111 (199).

\bibitem{RAY1} T.S. Ray, in {\it Artificial Life II}: Proceedings
  of the 2nd International Workshop on the Synthesis and Simulation of
  Living Systems, eds. Langton, C.G., Taylor, C., Farmer J.D., and
Rasmussen S. (Addison Wesley, Redwood City, 1992), p. 371.

\bibitem{AB1} C. Adami, and C.T. Brown, in {\it Artificial Life
    IV: Proceedings of the 4th International Workshop on the Synthesis
    and Simulation of Living Systems}, eds. Brooks, R. A. \& Maes, P.
  (MIT Press, Cambridge, MA), p. 377 (1994).

\bibitem{adami98} C. Adami, {\it Introduction to Artificial
    Life} (Springer, New York, 1998).

\bibitem{CHU1} J. Chu and C. Adami. Propagation of information in
populations of self-replicating code, Proc. of "Artificial Life V",
Nara (Japan), May 16-18, 1996, C.G. Langton and T. Shimohara, eds.,
(MIT Press, Cambridge, MA, 1997), p. 462-469.


\bibitem{adami98a}C. Adami, R. Yirdaw, and R. Seki, Critical exponent
of species-size distribution in evolution, Proc. of "Artificial Life
VI" Los Angeles, June 27-29, 1998; C. Adami, R. Belew, H. Kitano, and
C. Taylor, eds., (MIT Press, Cambridge, MA, 1998), p. 221-227.

\bibitem{ofria99a} C. Ofria, C. Adami, T.C. Collier, and
G. Hsu. Evolution of differentiated expression patterns in digital
organisms. {\it Lect. Notes Artif. Intell.} {\bf 1674}:129-138 (1999).

\bibitem{ofria99b} C. Ofria and C. Adami. Evolution of genetic
organization in digital organisms. Proc. of DIMACS Workshop on
Evolution as Computation, Jan 11-12, Princeton University,
L. Landweber and E.  Winfree, eds., Springer-Verlag, N.Y. (1999)
p. 167.

\bibitem{Lenskietal99} R.E. Lenski, C. Ofria, T.C. Collier, and
C. Adami, Genome complexity, robustness and genetic interactions in
digital organisms, {\it Nature} {\bf 400}:661--664 (1999).

\bibitem{adami00}C. Adami, C. Ofria, and T.C. Collier, Evolution of
biological complexity, {\it Proc. Nat. Acad. Sci. USA} {\bf
97}:4463-4468 (1999).

\bibitem{wagenaar00}D. Wagenaar and C. Adami, Influence of chance,
history and adaptation on evolution in {\it Digitalia}, Proc. of
Artificial Life VII, M.A. Bedau, J.S. McCaskill, N.H. Packard, and
S. Rasmussen, eds. (MIT Press, 2000), p. 216-220.

\bibitem{ieee2001} C. Ofria, C. Adami, and T.C. Collier. Design of
evolvable computer languages. {\it IEEE Trans. Evol. Comp.} {\bf 6}:420--424
(2002).

\bibitem{hutch61}G. E. Hutchinson, The paradox of the plankton, {\it
Am. Nat.} {\bf 95}:137--145 (1961).

\bibitem{huis00} J. Huisman and F. J. Weissing, Biodiversity of
plankton by species oscillation and chaos, {\it Nature} {\bf
402}:407--410 (1999).

\end{thebibliography}
\end{document}